\documentclass[12pt,preprint]{aastex}
\begin{document}
\title{Non-WKB Models of the FIP Effect: Implications for Solar Coronal
Heating and the Coronal Helium and Neon Abundances}


\author{J. Martin Laming}
\affil{Space Science Division, Naval Research Laboratory Code 7674L,
Washington, D.C. 20375} 

\begin{abstract}
We revisit in more detail a model for element abundance fractionation
in the solar chromosphere, that gives rise to the ``FIP Effect'' in the solar
corona and wind. Elements with first ionization potential below about 10 eV, i.e. those
that are predominantly ionized in the chromosphere, are
enriched in the corona by a factor 3-4. We model the propagation of Alfv\'en waves
through the chromosphere using a non-WKB treatment, and evaluate the ponderomotive
force associated with these waves. Under solar conditions, this is generally
pointed upwards in the chromosphere, and enhances the abundance of chromospheric
ions in the corona. Our new approach captures the essentials of the solar coronal
abundance anomalies, including the depletion of He relative to H, and also the
putative depletion of Ne, recently discussed in the literature. We also argue that
the FIP effect provides the strongest evidence to date for energy fluxes
of Alfv\'en waves sufficient to heat the corona. However it appears that these waves
must also be generated in the corona, in order to preserve the rather regular
fractionation pattern without strong variations from loop to loop
observed in the solar corona and slow speed solar wind.
\end{abstract}


\section{Introduction}

Since about 1985 it has been recognized that the composition of the
solar corona and the photosphere are not the same.  In the corona,
the ratios of abundances of elements with first ionization
potentials (FIP) less than about 10 eV relative to abundances of
elements with a FIP greater than about 10 eV are about a factor of
3-4 times higher than in the photosphere.  Those elements with a FIP
greater than 10 eV appear to have a photospheric composition in
general (with respect to hydrogen) in the corona, and the low FIP
elements are enhanced in abundance there. This fractionation has
recently been explained \citep{laming04} as being due to the
ponderomotive force in the chromosphere from Alfv\'en waves. This is
usually directed upwards, and acts on chromospheric ions, but not
neutrals. Elements that are predominantly ionized in the
chromosphere (low FIP elements like Al, Mg, Si, Ca, and Fe) can be
enhanced in abundance as they flow up to the corona, whereas high
FIP elements such as C, N, O, Ne, and Ar that are largely neutral
appear essentially unaffected. The abundance of sulfur (FIP = 10.36
eV) is between photospheric and coronal.

It was recognized very early that the most plausible site for FIP
fractionation was the chromosphere, where low FIP elements are
generally ionized and where high FIP elements are at least partially
neutral.  Henoux (1995, 1998) reviewed the early models. Those which
rely on ion-neutral separation by diffusion in magnetic fields, in
temperature gradients, or in upward plasma flows suffer from
problems of the speed of the process (diffusion is inherently slow)
or the choice of boundary conditions.  Realistic FIP effect models
must include some form of external force that acts upon the plasma
ions but not upon the neutrals.  The first effort along these lines
(Antiochos 1994) considered the cross-B thermoelectric electric
field associated with the downward heat flux carried by electrons
which gives rise to chromospheric evaporation.  This draws ions into
the flux tube, enhancing their abundance over neutrals.  The absence
of a FIP effect in coronal holes arises naturally, but in coronal
regions where FIP fractionation occurs, a mass dependence is
predicted, which is not observed.

Henoux \& Somov (1992) proposed that cross-B pressure gradients
arising in current carrying loops could enhance the ion abundance by
a ``pinch'' force.  Azimuthal motions of the partially ionized
photosphere at flux tube boundaries generate a system of currents
flowing in opposite directions, such that the azimuthal component of
the field vanishes at infinity.  Details of the fractionation (mass
dependence, degree, etc) remain to be worked out, but it is thought
to begin just above the temperature minimum region at about 4000 K,
and continue until temperatures where all elements are ionized. More
recent suggestions have been that chromospheric ions, but not
neutrals, are heated by either reconnection events (Arge \& Mullan
1998), or by waves that can penetrate down to loop footpoints from
the corona (Schwadron, Fisk \& Zurbuchen 1999). Strengths and
weaknesses of these two models are discussed in some detail by
Laming (2004).  For the time being we comment that although both
models seem capable of producing mass independent fractionation of
about the right degree, they, in common with the others mentioned
above, only predict positive FIP effects.  In these models there is
no possibility of an ``inverse FIP'' effect such as seen in the
coronae of active stars (see e.g. Feldman \& Laming 2000, Laming
2004 and references therein).

This consideration led Laming (2004) to consider the action of
ponderomotive forces due to Alfv\'en waves propagating up through
the chromosphere and either transmitting into or being reflected
from coronal loops.  This force can in principle be either upward or
downward and is given approximately by (see derivation in Appendix
A)
\begin{equation}
F={q^2\over 4m\left(\Omega ^2-\omega ^2\right)}{\partial\delta E_{\perp}^2\over\partial
z}
\end{equation}
where $\Omega$ and $\omega$ are the particle cyclotron frequency and the wave
frequency respectively, $q$ and $m$ the charge and mass, and $E_{\perp}$ is the
wave peak transverse electric field. The
dependence on the Alfv\'en speed, $V_A$, means that the
ponderomotive force is usually strongest at the top of the
chromosphere. The ponderomotive acceleration, $F/m$, is independent
of the ion mass, leading to the essentially mass independent
fractionation that is observed. Using an analytic model solar loop
from Hollweg (1984), upward ponderomotive forces on ions are much
more common in solar conditions, and for typical density gradients
in the chromosphere can be larger than the gravitational force
downward. The magnitude of the FIP fractionation is dictated by the
resonant properties of the coronal loop and the corresponding wave
energy density.  Large loops with resonant frequencies similar to
the chromospheric period of 200-300 seconds admit strong Alfv\'en
wave fluxes with correspondingly large FIP fractionation.  Open
field lines, or coronal holes, with formally infinite wave periods
or unresolved fine structures (i.e. the chromosphere and lower
transition region away from coronal loop footpoints, Feldman 1983,
1987) with much shorter wave periods do not have much Alfv\'en wave
transmission into them and so would have very low FIP fractionation,
as observed. Using model chromospheres from Vernazza, Avrett, \&
Loeser (1981), most of the FIP fractionation is found to occur at
the top of the chromosphere at altitudes with strong density
gradients near the plateau regions, where low FIP elements are
essentially completely ionized and high FIP elements are typically
at least 50\% ionized. The Laming (2004) model thus comes about as a
natural extension of existing work on Alfv\'en wave propagation in
the solar atmosphere with essentially no extra physics required.

In this paper we revisit the \citet{laming04} model using a
numerical treatment of Alfv\'en wave propagation in a coronal loop
rooted in the chromosphere at each footpoint. Section 2 describes
this model and the improvements over \citet{laming04} with a series
of illustrative examples. Section 3 gives a more complete tabulation
of the FIP fractionation expected in a variety of elements, and
section 4 gives some discussion of the implications of the new
models, both for the solar coronal abundances of helium and neon,
and for MHD wave origin and propagation in the solar atmosphere,
before section 5 concludes.

\section{Ponderomotive Driving of the FIP Effect}
\subsection{Introduction and Formalism}
Laming (2004) used a WKB approximation to treat the case of strong
transmission of Alfv\'en waves into a coronal loop, and hence
evaluate the FIP fractionation. Here we have extend this work
to a non-WKB treatment. The procedure follows that described in
detail by Cranmer \& van Ballegooijen (2005), but applied to closed
rather than open magnetic field structures. The transport equations
are (see Appendix B for a derivation)
\begin{equation} {\partial I_{\pm}\over\partial t}+\left(u\pm
V_A\right){\partial I_{\pm}\over\partial z}= \left(u\pm
V_A\right)\left({I_{\pm}\over 4H_D}+{I_{\mp}\over 2H_A}\right),
\end{equation}
where $I_{\pm}=\delta v\pm \delta B/\sqrt{4\pi\rho }$ are the
Els\"asser variables for inward and outward propagating Alfv\'en
waves respectively. The Alfv\'en speed is $V_A$, the upward flow
speed in $u$ and the density is $\rho $. The signed scale heights
are $H_D=\rho /\left(\partial\rho /\partial z\right)$ and
$H_A=V_A/\left(\partial V_A/\partial z\right)$. In the solar
chromosphere and in closed loops we may take $u << V_A$.

\begin{figure}
\epsscale{0.75} \plotone{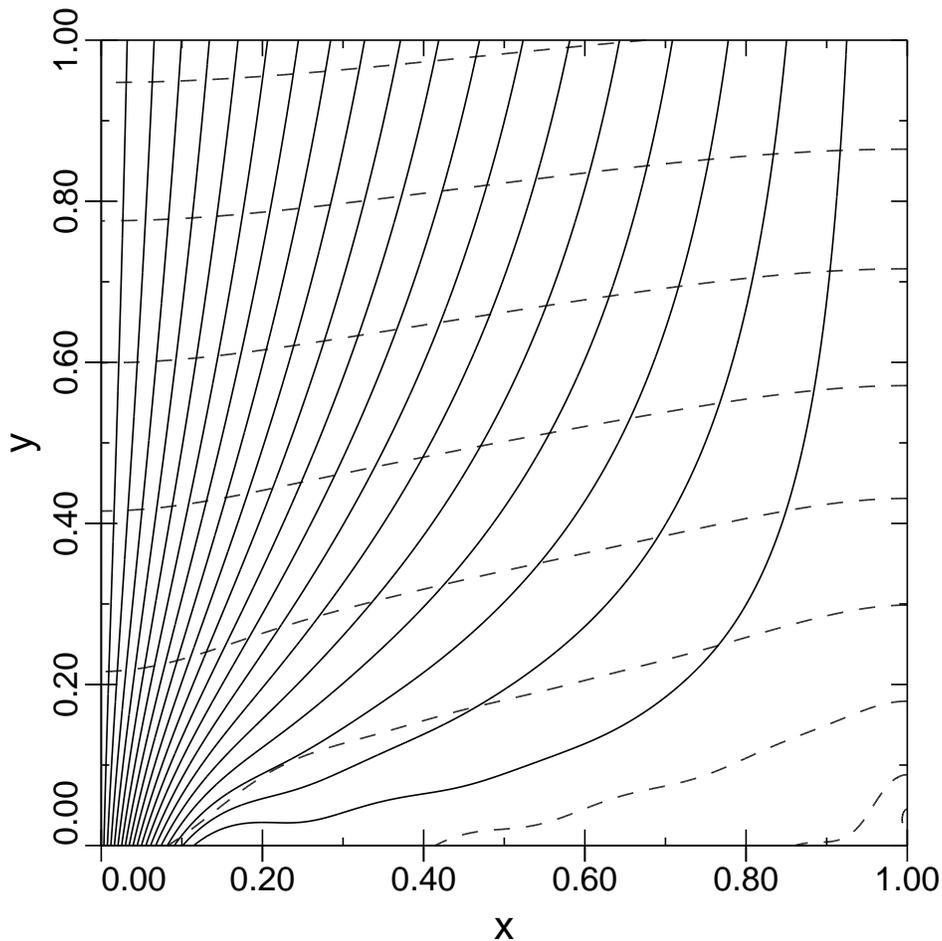} \figcaption[f1.eps]{Force free
magnetic field, computed from Athay (1981),  from the center of a
network segment ($x=0$) to the center of a supergranule cell
($x=1$). We take $x=1$ to represent 1000 km, and y=0 to represent an
altitude of 500 km above the photosphere. The solid lines represent
magnetic lines of force, and dashed lines are logarithmically
spaced contours of the Alfv\'en
speed, assuming the density falls off exponentially with height. FIP fractionation
in this work occurs towards the top of the chromosphere, where the magnetic
field is nearly parallel.
\label{fig1}} \end{figure}

\begin{figure}
\epsscale{0.75} \plotone{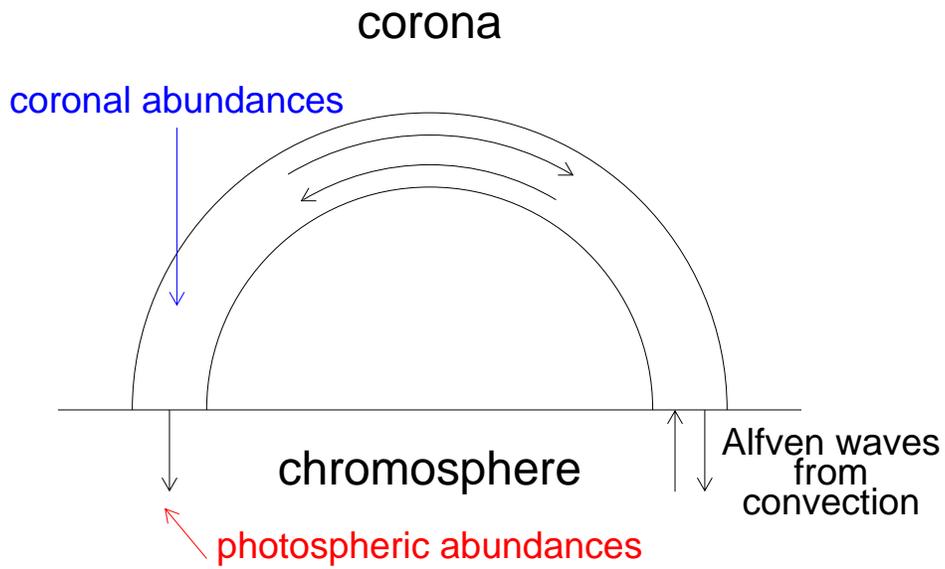} \figcaption[f2.eps]{Cartoon
illustrating the model. Alfv\'en waves are incident on the coronal
loop from below on the right hand side. Waves are either transmitted
into the loop or reflected back down again. Waves in the coronal
loop bounce back and forth, with some leakage at each footpoint. The
magnetic field is taken to be uniform in the coronal section of the
loop (illustrated), while it varies according to Figure 1 within the
chromosphere (not shown on this figure).
\label{fig2}}
\end{figure}

We use the same chromospheric model as before in Laming (2004), but
this time also embedding it in a 2D force-free magnetic field
computed from formulae given in \citet{athay81}, and shown in Figure
\ref{fig1}. We take a scale here of 1 unit to 1,000 km, and place
the bottom of the plot at 500km altitude in the chromosphere. The
region where the plasma beta (ratio of gas pressure to magnetic
pressure) equals unity is taken at 650 km altitude, or at $y=0.15$
in the figure. This region is where upcoming acoustic waves from the
convection zone can convert to Alfv\'en waves by mode or parametric
conversion, or where downgoing Alfv\'en waves can convert back to
acoustic waves. We adopt an altitude of 650 km as the boundary of
our simulations where Alfv\'en waves are launched upwards. We
consider a loop model similar to that of \citet{hollweg84}, the
coronal portion of which is illustrated in Figure \ref{fig2}. Waves
from beneath impinge on the right hand side chromosphere-corona
boundary and are either reflected back down, or transmitted into the
loop. Waves in the loop section bounce back and forth, with a small
probability of leaking back into the chromosphere at either end. The
magnetic field in the coronal loop section is uniform, and is simply
the extension of chromospheric force-free field into the corona. In
the current work, the non-uniformity of the magnetic field in the
low chromosphere has no effect on the FIP effect, since the fractionation
in this work appears towards the top of the chromosphere where the magnetic field
is almost parallel. The loop
density is similarly extrapolated, but with a density scale height
taken here to be equal to the loop length. All models presented here
assume waves propagating up the $y$-axis at $x=0$.

Equations (2) are integrated from a starting point in the left hand
side chromosphere (hereafter chromosphere ``A'') where Alfv'en waves
leak down into the chromosphere, back through the corona to the
right hand side (chromosphere ``B'') where waves are fed up from
below. In this way the reflection and transmission of Alfv\'en waves at
the loop footpoints and elsewhere is naturally reconstructed.
The velocity and magnetic field perturbations are
calculated from
\begin{eqnarray}
\delta v&={I_++I_-\over 2}\cr
{\delta B\over\sqrt{4\pi\rho}}&={I_+-I_-\over 2}.
\end{eqnarray}
The wave energy density and positive and negative going energy fluxes are
\begin{eqnarray}
U&={\rho\delta v^2\over 2} +{\delta B ^2\over 8\pi}={\rho\over 4}
\left(I_+^2 + I_-^2\right)\cr
F_+&={\rho\over 4}I_+^2V_A\cr
F_-&={\rho\over 4}I_-^2V_A,
\end{eqnarray}
and the wave peak electric field appearing in equation (1) is
\begin{equation}
\delta E_{\perp}^2={B^2\over 2c^2}\left(I_+^2+I_-^2\right).
\end{equation}

Given the ponderomotive acceleration from equation (1), the FIP fractionations are
calculated in a similar manner to \citet{laming04}, but with one
important modification. In the momentum equations (10) and (11) of
\citet{laming04}, we add the motion in the wave to the ion and
neutral partial pressures, so that $P_{s,i,n}=\left(k_{\rm B}T/m_s
+v_{turb}^2+v_{wave}^2\right)\rho _{s,i,n}/2$ where the first two
terms in parentheses represent the ion thermal velocity and the
microturbulent velocity in the model chromosphere, and the third
term represents the motion of the ion in the Alfv\'en wave. In true
collisionless plasma, neutrals would not respond to the wave.
However the solar chromosphere is sufficiently collisional that
neutrals move with the ions in the wave motion
\citep[e.g.][]{vranjes08} for wave frequencies well below the charge
exchange rate that couples neutrals and ions, and so neutrals
require the same form for their partial pressure as the ions.
Following the derivation through, writing $\nu _{s,n}\partial
P_{s,i}/\partial z + \nu _{s,i}\partial P_{s,n}/\partial z$ from
equations (10) and (11) in \citet{laming04} (with the ponderomotive
term in $\partial\rho _{s,i}/\partial z\rightarrow 0$, i.e. $b=0$)
we find
\begin{equation}
{\partial\over\partial z}\left[{\rho _s\over 2}\left({k_{\rm
B}T\over m_s}+v_{turb}^2+v_{wave}^2\right)\right]+\rho_s\left[g+\nu
_{eff}\left(u_s-u\right)\right]+\rho _sa\xi _s\nu _{eff}/\nu _{si}=0
\end{equation}
where $\nu_{eff}=\nu _{s,i}\nu _{s,n}/\left[\xi _s\nu
_{s,n}+\left(1-\xi _s\right)\nu _{s,i}\right]$, with $\xi _s$ being
the ionization fraction of element $s$, and $\nu _{s,i}$ and $\nu
_{s,n}$ the collision rates of ions and neutrals respectively of
element $s$ with the ambient gas, and $a$ the ponderomotive
acceleration. This leads to
\begin{equation}
{\rho _s\left(z_u\right)\over\rho _s\left(z_l\right)}=
{v_s\left(z_l\right)^2\over v_s\left(z_u\right)^2} \exp\left(2\int
_{z_l}^{z_u}{-g-\nu _{eff}\left(u_s-u\right)+\xi _sa\nu _{eff}/\nu
_{s,i}\over v_s^2}dz\right)
\end{equation}
where $v_s^2=kT/m_s +v_{turb}^2 +v_{wave}^2$.
We argue that in the absence of the ponderomotive
acceleration $a$, the effect of turbulence should be to fully mix
the element composition.
Thus we choose $u_s$ in equation (7) such that
\begin{equation}
{v_s\left(z_l\right)^2\over v_s\left(z_u\right)^2} \exp\left(2\int
_{z_l}^{z_u}{-g-\nu _{eff}\left(u_s-u\right)\over v_s^2}dz\right)=1
\end{equation}
to yield the fractionation by the ponderomotive force as
\begin{equation}
{\rho _s\left(z_u\right)\over\rho
_s\left(z_l\right)}=\exp\left(2\int _{z_l}^{z_u}{\xi _sa\nu
_{eff}/\nu _{s,i}\over v_s^2}dz\right).
\end{equation}

As mentioned previously \citep{laming04}, the solar chromosphere is
undoubtedly more active and dynamic than represented by equations
(6-9). However this choice allows us to model a chromosphere which
in the absence of Alfv\'en waves is completely mixed, presumably by
hydrodynamic turbulence, and upon which the ponderomotive force acts
to selectively accelerate chromospheric ions. Chromospheric simulations excluding
turbulence find huge (and unobserved) variations in
various coronal element abundances due to ambipolar and thermal diffusion
\citep{killie07}. It would appear that such abundance variations are
unavoidable in the situation that the chromosphere remains undisturbed for
a sufficient length of time \citep[days to weeks in the case of][]{killie07}.
We argue
that hydrodynamic turbulence acts on timescales much shorter than this (but still
longer than that required to establish the FIP effect) to leave a fully
mixed chromosphere in the absence of ponderomotive forces. We model the
ponderomotive acceleration as acting only on the chromospheric ions,
since the ponderomotive acceleration divided by the flow velocity
$a/u \sim 10 - 1000$ s$^{-1}$ is much greater than the charge
exchange rate (of order 1 s$^{-1}$). This rate should also be sufficiently
greater than the turbulent mixing rate discussed above.

We also use a more recent chromospheric model \citep[model C7
in][]{avrett08}, introduced as an update to the older VALC model
\citep{vernazza81} used previously in \citet{laming04}. We continue
to evaluate photoionization rates using incident spectra based on
\citet{vernazza78} with the extensions and modifications outlined in
\citet{laming04}. In most cases, the ``very active region'' spectrum
is used, being the most consistent with the underlying chromospheric
model.

\subsection{A Loop On Resonance}
\begin{figure}[t]
\epsscale{1.0} \plotone{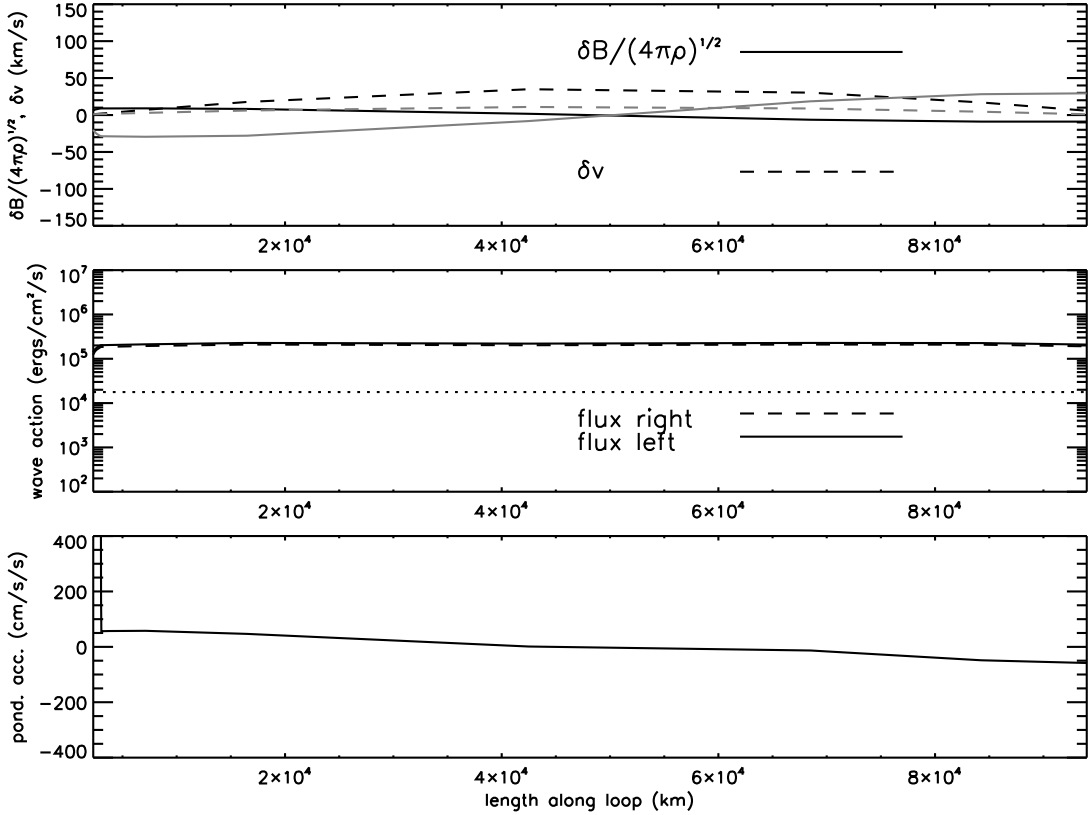} \figcaption[f3.ps]{Coronal section of
loop, length 100,000 km, magnetic field 9.9 G, (half wavelength
long) showing from top: Els\"asser variables in km s$^{-1}$ ($\delta
B/\sqrt{4\pi\rho}$ solid lines, $\delta v$ dashed lines), with black
lines for real parts and gray lines for imaginary parts. The loop is
approximately half a wavelength long. Middle; wave energy fluxes in
ergs cm$^{-2}$ s$^{-1}$, the thin solid line shows the difference in
energy fluxes divided by the magnetic field strength and should be a
horizontal line if energy is properly conserved. Bottom, the
ponderomotive acceleration in cm s$^{-2}$. Positive acceleration means
positive along the $z$ axis, which is upwards pointing near $z=0$ and
downwards near $z=100,000$.\label{fig3}}
\end{figure}

\begin{figure}[t]
\epsscale{1.0} \plotone{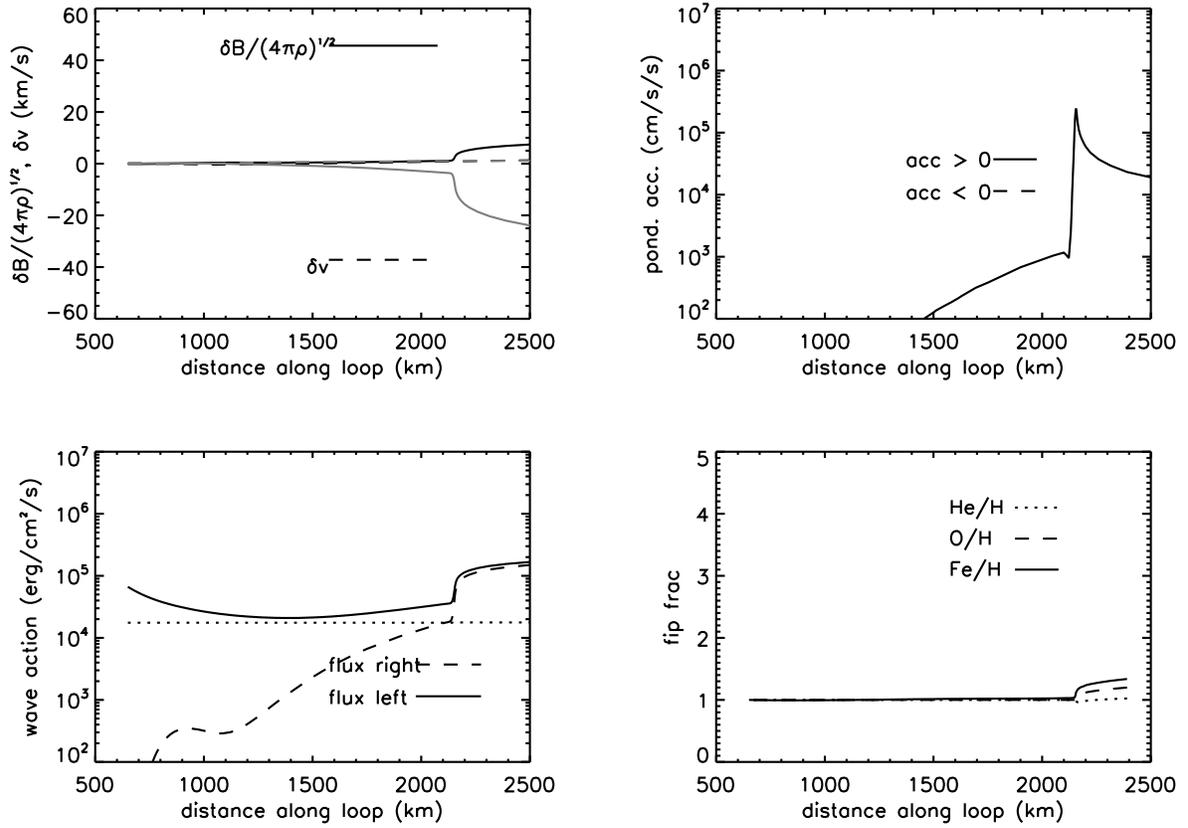} \figcaption[f4.ps]{Same as figure 3
giving the first three panels for the left hand chromosphere ``A'', where
waves leak down from the corona. The extra bottom right panel shows
the FIP fractionation for Fe, O, and He, relative to H.\label{fig4}}
\end{figure}

\begin{figure}[t]
\epsscale{1.0} \plotone{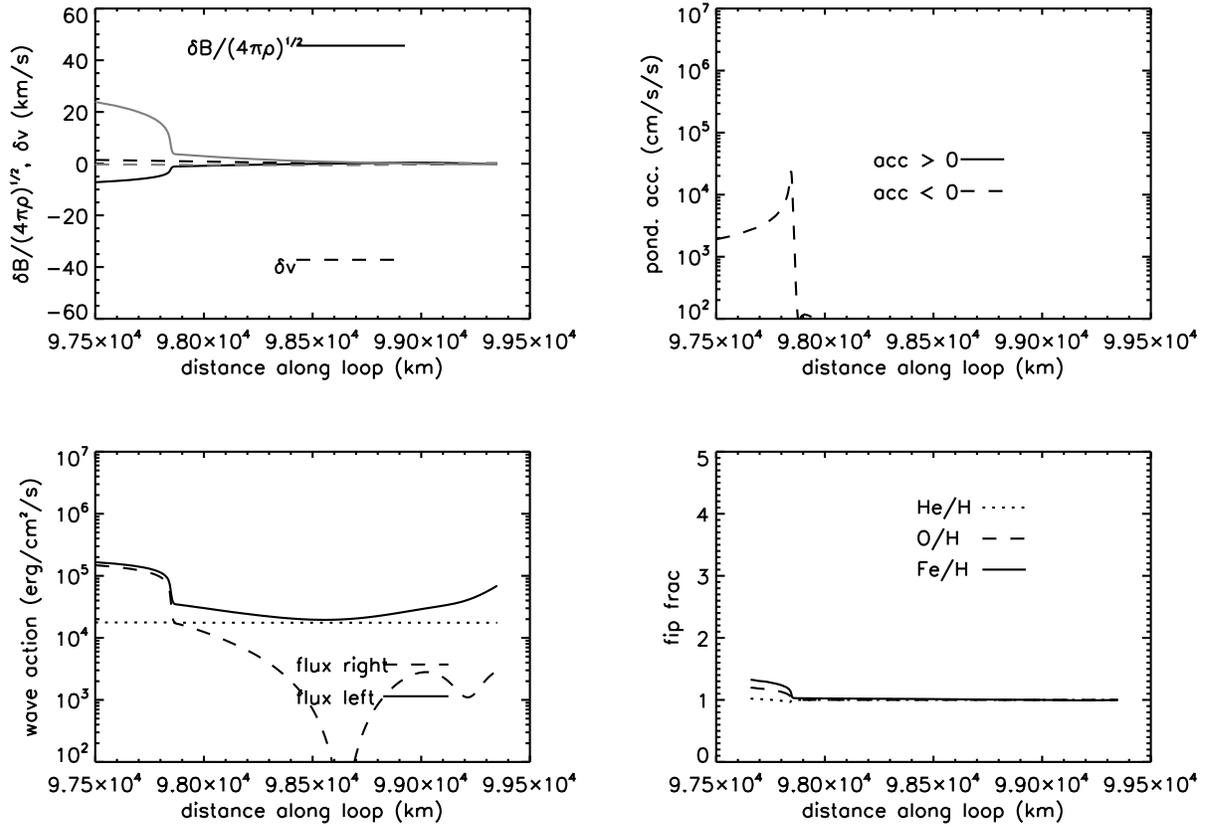} \figcaption[f5.ps]{Same as figure 4
for the right hand side chromosphere ``B'', where Alfv\'en waves are
launched up from the convection zone.\label{fig5}}
\end{figure}

In the following, Figures \ref{fig3}, \ref{fig4}, and \ref{fig5}
illustrate the solutions for a loop 100,000 km long, with coronal
magnetic field $B=9.9$ G. This yields a wavelength for 3 minute
period waves approximately the same as twice the loop length, and
therefore such waves can be transmitted into the coronal loop
section from the chromosphere. We concentrate on 3 minute waves since unlike
5 minutes waves, these require no special conditions to propagate up into
the corona \citep{depontieu05}. Figure \ref{fig3} shows the coronal
section of the loop. From top to bottom the three panels give the
amplitudes of $\delta v$ and $\delta B/\sqrt{4\pi\rho }$ in units of km s$^{-1}$. Real
and imaginary parts are given as black and gray lines respectively,
with $\delta B/\sqrt{4\pi\rho }$ and $\delta v$ given by solid and
dashed line respectively. The wave amplitude has been chosen to be $\sim 30$
km s$^{-1}$ in the corona, giving a typical spectral line FWHM consistent with
observations \citep[e.g.][]{mcintosh08}. The second panel gives the wave actions
(energy fluxes) for the left going (solid) and right going (dashed)
lines, and their difference divided by the magnetic field as a
dotted line, in arbitrary units. This last quantity should be a
straight horizontal line if energy is properly conserved in the
calculation. The third panel gives the ponderomotive acceleration,
in cm s$^{-2}$. Throughout the coronal section of the loop, it is
significantly lower than the gravitational acceleration. Solid lines
indicate positive, i.e. right going, and dashed lines indicate left
going accelerations. The oscillation amplitude has been chosen to give
mass motions within observational constraints \citep{chae98,mcintosh08},
as measured from line profiles.

Figure \ref{fig4} shows the same three plots for the left hand side
chromosphere ``A'', where waves leak down from the corona, together
with a fourth panel showing the degree of fractionation for the
abundance ratios Fe/H, O/H, and He/H. The ponderomotive acceleration
in the chromosphere is much larger than in the corona, especially
towards the top. The most significant fractionation occurs here,
increasing Fe/H in this case by a factor of 1.4 over photospheric
values, O/H by a factor of around 1.25, with He/H remaining nearly
unchanged.

Finally Figure \ref{fig5} shows the same four panels for the
chromosphere ``B'' on the right hand side, where the upgoing
Alfv\'en waves originate. The ponderomotive acceleration is still
pointed upwards (though is negative in the coordinate system used
here), giving the same FIP fractionations as before. In exact
resonance, the chromospheric ponderomotive force behaves the same as
at the opposite footpoint already shown in Figure \ref{fig4}.

\subsection{A Loop Off Resonance}
\begin{figure}[t]
\epsscale{1.0} \plotone{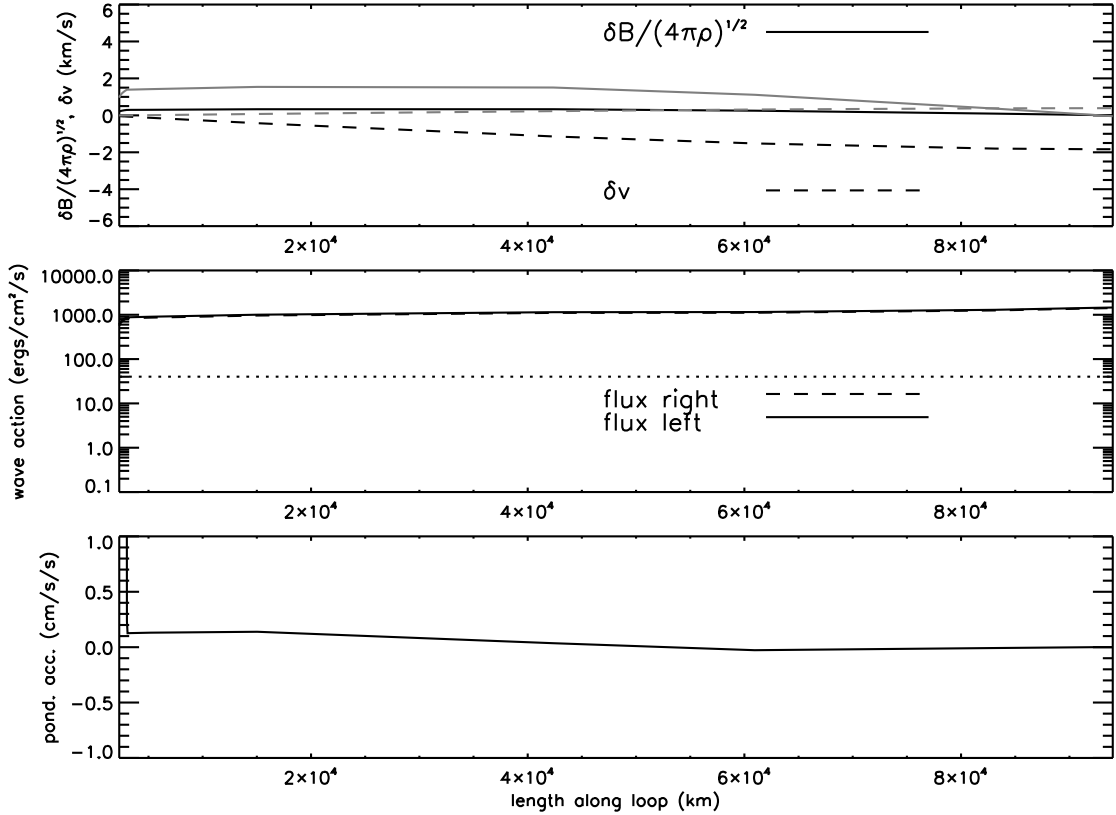} \figcaption[f6.ps]{Coronal section of
loop, length 100,000 km, magnetic field 19.8 G, showing from top:
Els\"asser variables in km s$^{-1}$ ($\delta B/\sqrt{4\pi\rho}$
solid lines, $\delta v$ dashed lines), with black lines for real
parts and gray lines for imaginary parts. The loop is now a quarter
wavelength long and reflects most Alfv\'en waves incident from
below, and consequently has much smaller nonthermal motions than the
previous resonant case. Middle; wave energy fluxes in ergs cm$^{-2}$
s$^{-1}$. Bottom, the ponderomotive acceleration in cm s$^{-2}$.
\label{fig6}}
\end{figure}

\begin{figure}[t]
\epsscale{1.0} \plotone{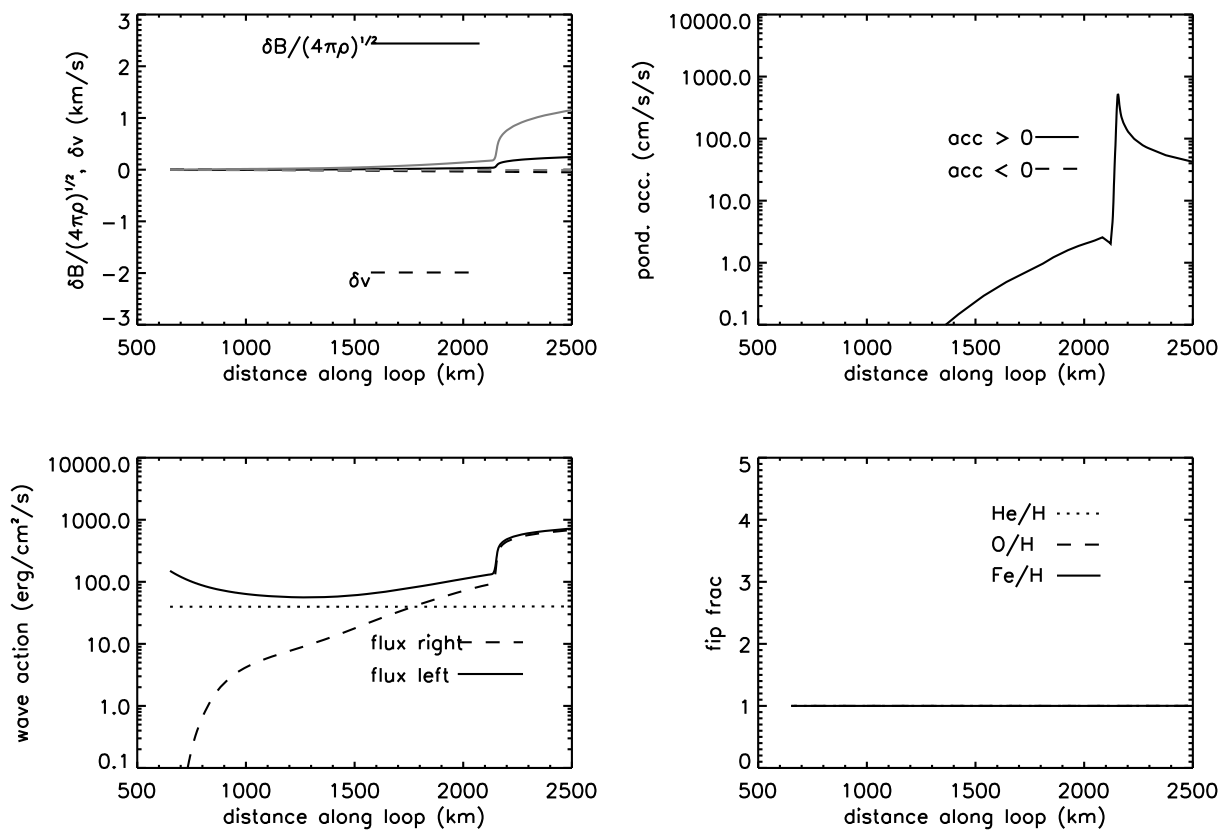} \figcaption[f7.ps]{Same as figure 6
giving the first three panels for the left hand chromosphere ``A'', where
waves leak down from the corona. The extra bottom right panel shows
the FIP fractionation for Fe, O, and He, relative to H. In the case of a loop
off resonance, no waves are transmitted through to chromosphere ``A'', and
no fractionation occurs.\label{fig7}}
\end{figure}

\begin{figure}[t]
\epsscale{1.0} \plotone{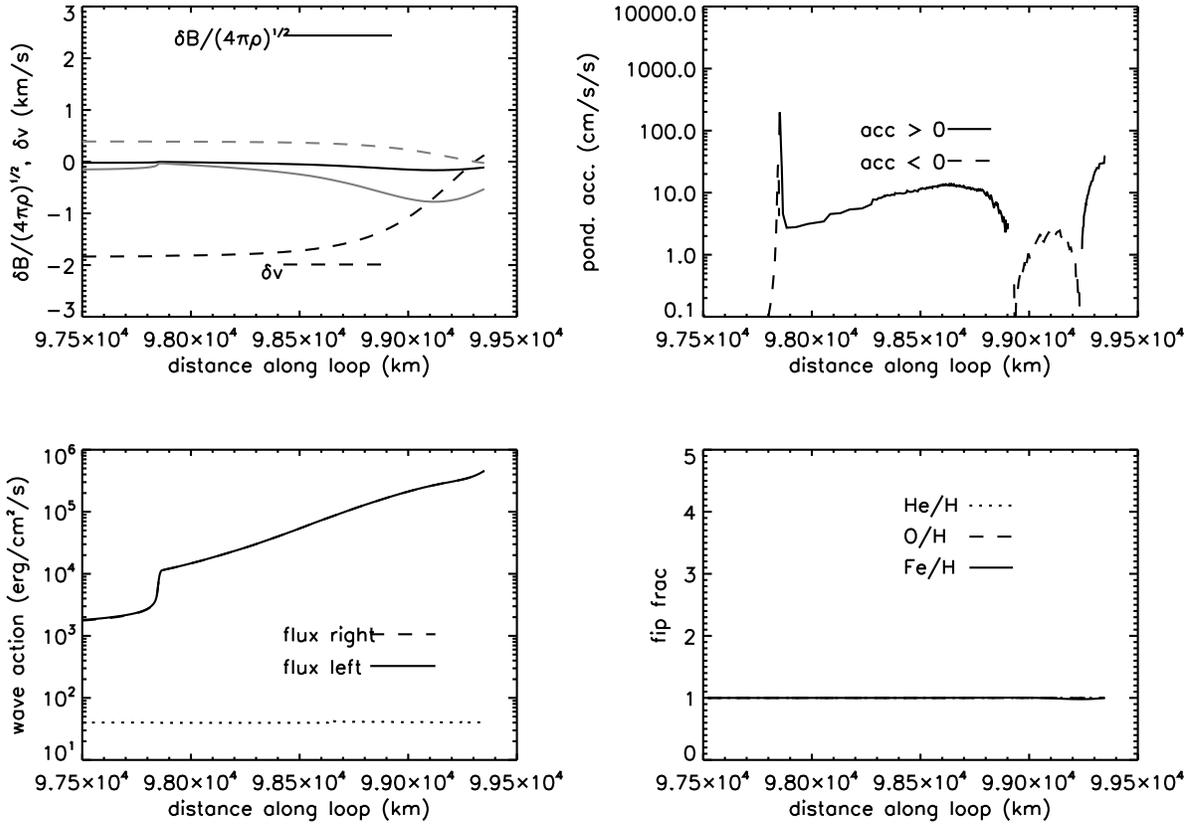} \figcaption[f8.ps]{Same as figure 7
for the right hand side chromosphere ``B'', where Alfv\'en waves are
launched up from the convection zone. Almost complete reflection of Alfv\'en
waves occurs from the loop footpoint, leading to no fractionation.\label{fig8}}
\end{figure}

Figures \ref{fig6}, \ref{fig7} and \ref{fig8} show the same
variables as before, but for a loop 100,000 km long and with
magnetic field $B=19.8$ G. Now the loop is a quarter wavelength
long, and almost complete reflection of the incident Alfv\'en waves
on the right hand side takes place. The simulation has been
normalized so that the incident Alfv\'en wave flux coming up from
the chromosphere is about the same as for the on resonance case.
Figure \ref{fig6} shows that the coronal loop oscillation is now
much weaker than before, by about a factor 20. In the left hand
chromosphere ``A'' (Figure \ref{fig7}) negligible Alfv\'en wave flux
leaks through and no FIP fractionation occurs. In the right hand
chromosphere ``B'' (Figure \ref{fig8}), the behavior is quite
different to the previous case. The ponderomotive force is now {\it
downwards} pointing for most heights in the chromosphere, and it is
still very small, also giving essentially no FIP fractionation. The
downward directed ponderomotive force might be of interest in cases
where the turbulence is stronger. As reviewed in \citet{laming04},
the coronae of various active stars exhibit an inverse FIP effect,
where the low FIPs are depleted in the corona instead of being
enhanced. The reversal of the ponderomotive force under these
conditions is a plausible mechanism for such abundance anomalies.

\subsection{Loops with Stronger Turbulence}
\begin{figure}[t]
\epsscale{1.0} \plotone{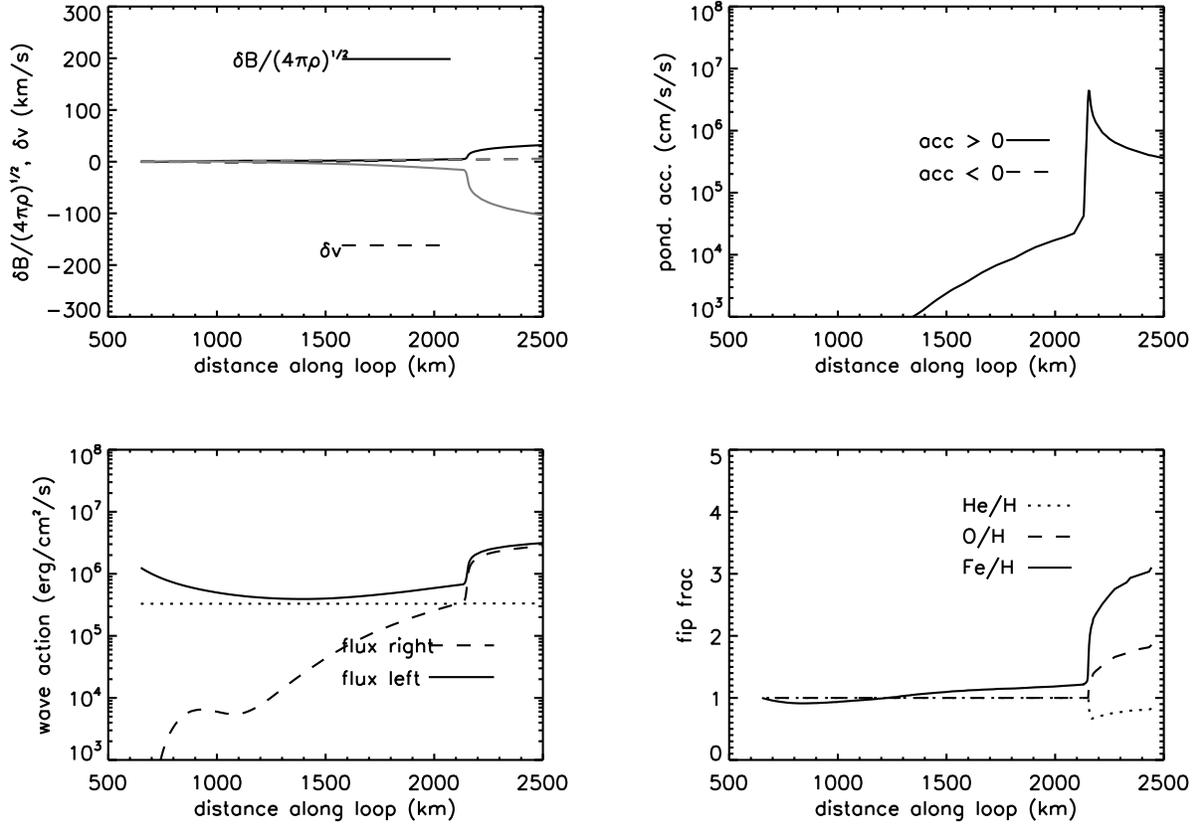} \figcaption[f9.ps]{Same as figure 4 (on
resonance case)
giving the first three panels for the left hand chromosphere ``A'', where
waves leak down from the corona. The extra bottom right panel shows
the FIP fractionation for Fe, O, and He, relative to H. The wave
energy flux has been increased by a factor 20, leading to stronger
coronal nonthermal motions, and stronger fractionation. Helium is now
depleted in the corona relative to the chromosphere.\label{fig9}}
\end{figure}

\begin{figure}[t]
\epsscale{1.0} \plotone{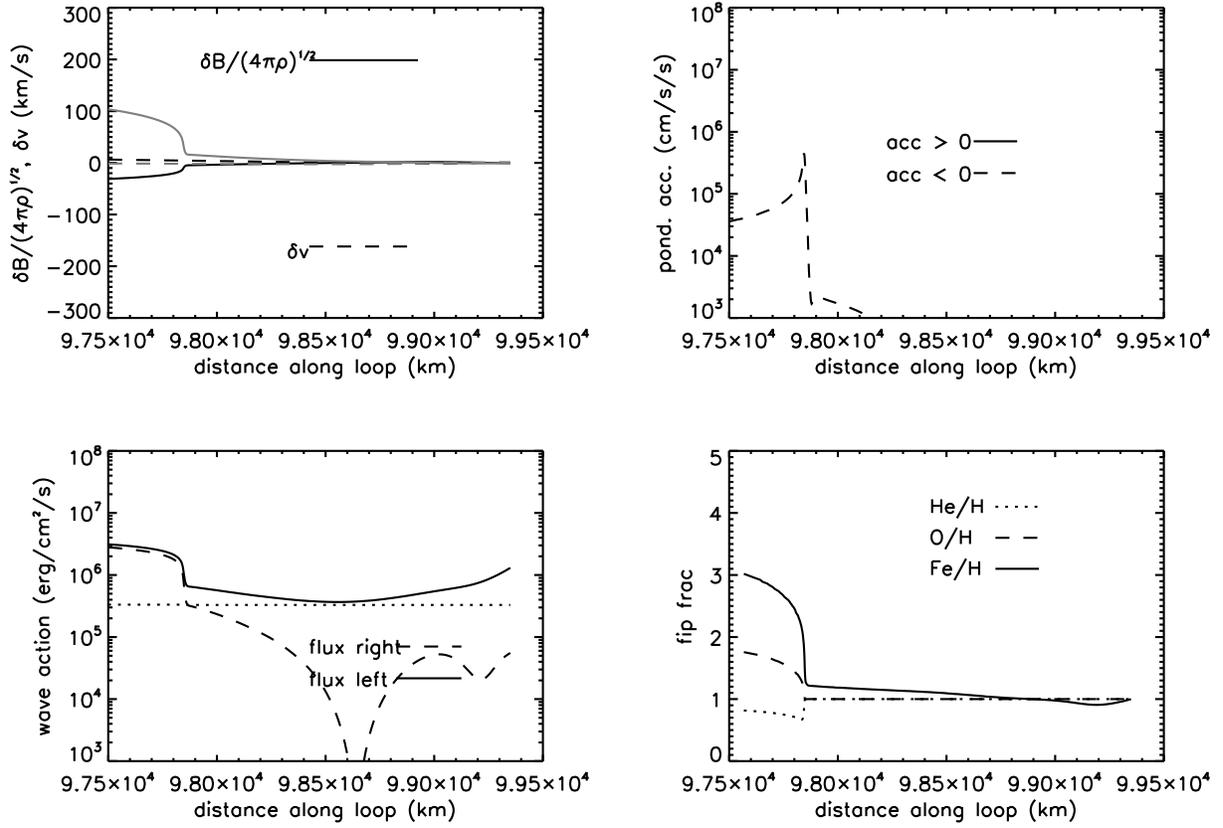} \figcaption[f10.ps]{Same as figure 5
(on resonance case)
for the right hand side chromosphere ``B'', where Alfv\'en waves are
launched up from the convection zone. The wave energy flux has been
increased by a factor 20, leading to stronger coronal nonthermal
motions, and stronger fractionation.\label{fig10}}
\end{figure}

The first case above was designed to give a coronal nonthermal mass
motion within observational limits, i.e. a root mean square $\delta
v\simeq 30$ km s$^{-1}$. The upgoing energy flux of Alfv\'en waves
at the loop footpoint is $\sim 10^5$ ergs cm$^{-2}$ s$^{-1}$, and is
insufficient to power the coronal radiation power loss by one to two
orders of magnitude. In this subsection we consider the same loop as
in the first case, but with an Alfv\'en wave upward energy flux of
about $2\times 10^6$ ergs cm$^{-2}$ s$^{-1}$; sufficient to power
radiation from a 100,000 km loop with a density of $10^8-10^9$
cm$^{-3}$. The
predicted nonthermal mass motions in the corona are now unphysically
high, in excess of 100 km s$^{-1}$, unless we are able to argue that only
a small region of the corona oscillates with this speed. We discuss this
further in subsection 4.1. This is less of a problem in the transition region
where the ``classical'' transition region that connects a coronal
loop with the chromosphere has only recently been identified in
observations \citep{peter01}, being otherwise masked by ``unresolved
fine structures'' \citep{feldman83,feldman87}. \citet{peter01} in
fact observed nonthermal line broadening in what he interprets as
the ``classical'' transition region approaching the values modeled
in this section, and suggests that they arise from the passage of an
Alfv\'en wave with sufficient energy flux to heat the corona. In
this strong wave field, the behavior of the FIP fractionation is now
subtly different, as shown in Figures \ref{fig9} and \ref{fig10} for
the left and right hand side chromospheres (``A'' and ``B'')
respectively. On each side, Fe is somewhat more enhanced than in the
previous case, at 3 - 3.3. O has a FIP fractionation of about
1.7-1.8, similar to before, but He/H is now at about 0.8 of its
photospheric value.

This new behavior can be understood with reference to equation 5,
and the denominator in the integral, $v_s^2=kT/m_s +v_{turb}^2
+v_{wave}^2$. In the case that the first two
terms in $v_s^2$ dominate, i.e. weak Alfv\'en turbulence, all
elements (high FIPs as well as low FIPs) are fractionated positive
to H because H has the largest thermal velocity in the denominator.
When the Alfv\'enic velocity dominates, the fractionation changes
and is determined solely by the numerator in the integral. In this
case the element that stays neutral the longest, He, as expected,
has the lowest abundance in the corona, being depleted with respect
to H. This occurs because H experiences a stronger ponderomotive
enhancement. O/H is unchanged, again as expected because O and H
have very similar ionization potentials and their ionization
structures are locked by charge exchange reactions between them. Fe
remains fractionated with respect to H, by a similar amount as
before. The inclusion of the Alfv\'en turbulence in $v_s$ leads to a
natural saturation of the FIP effect, at about the level observed.
Thus for a wide range of turbulence levels, and FIP effect of around
3 should be expected.

The decrease in He/H is especially interesting. It might be relevant
to the He abundance in the solar wind, of around 4-5\%
\citep[e.g.][]{aellig01,kasper07} compared with a photospheric
abundance of 8\%, also seen in coronal holes and quiet solar corona
\citep{laming01,laming03}, and is discussed further below.

\section{More Realistic Examples}
We put three Alfv\'en waves with angular frequencies 0.025, 0.022,
and 0.016 rad s$^{-1}$, with relative intensities 1:0.5:0.25 in the
left hand chromosphere designed to match the network power spectrum
displayed in Figure 1 of \citet{muglach03}. The loop is 100,000 km
long as before, with a magnetic field of 7.1 G, which puts the 0.025
rad $s^{-1}$ on resonance. In the first case, FIP fractionations are
computed for the left hand chromosphere ``A'', using the very active
region spectrum of \citet{vernazza78}, and are given in Table 1.
This is the region where waves leak into the chromosphere from the
corona before being reflected back up again, and should give FIP
fractionation. The corresponding model is shown in Figure 11. There
are three important differences from the tabulation given previously
\citep[Table 2 in][]{laming04}. The first is that with increasing
wave energy flux, the FIP fractionations now appear to saturate at
levels corresponding to a fractionation of low FIP elements
overabundant with respect to high FIP elements by a factor of about
3, and does not increase without limit. This arises from the
inclusion of the term in $v_{wave}^2$ in the ion and neutral partial
pressures discussed above, and means that for a wide range of
turbulent energy densities, similar fractionated abundances should
result. The other new features, already mentioned briefly above, are
the depletion in the He abundance, and at higher energy fluxes also
the Ne abundance relative to H. These also stem from the
modification to the partial pressures.

These new calculations are compared in Table 1 with observations from
\citet{zurbuchen02}, \citet{bryans08} and \citet{giammanco08}. \citet{zurbuchen02} give
abundances measured in the slow speed solar wind during 1997/8 relative to
O, relative to photospheric abundances given by \citet{grevesse98}. \citet{bryans08}
give abundances observed spectroscopically in a region of quiet solar corona,
again tabulated relative to the photospheric composition of \citet{grevesse98},
with small modifications by \citet{feldman00}. With the exceptions of Mg and K,
the calculated abundances agree well with those observed for a wave energy flux
between one and four times that shown in Figure 11, both for the elements that
are depleted, like He and Ne, and for those enriched. We predict stronger
fractionation in Mg than is in fact observed, and stronger than in \citet{laming04}.
The reason for this has been tracked to the use of the newer chromospheric model
from \citet{avrett08}, where H retains a higher degree of ionization lower in the
chromosphere than in the previous VAL models. This then in turn renders the
ionization of Mg probably spuriously high because of charge transfer ionization
with the ambient protons. The difference in ionization fraction between 0.99 and
0.95 makes a considerable impact on the fractionations that result. Other low
FIP elements, Si, and Fe, do not have charge transfer ionization rates tabulated
by \citet{kingdon96}, and so are unaffected by this change. The cause of the discrepancy
for K is less clear. Like Na, K is very highly ionized throughout the chromosphere
due to its very low FIP, and should be expected to fractionate strongly, though
\citet{bryans08} do comment that their analysis only includes one line of K IX.

The results of the calculation for the right hand side chromosphere
``B'' are given in Figure 12. The loop
model chosen is resonant with the 0.025 angular frequency wave, and
this is the component transmitted into the corona. However the FIP
fractionation is significantly reduced by the presence of the other
wave frequencies which are reflected from the corona. In the right
hand chromosphere ``B'' the weaker components on the left are now
the strongest. This does not produce much change in the
ponderomotive acceleration, but increases the term $v_{wave}^2$ in
the denominator of the integrand in equation 6, thereby reducing the
fractionation. A wave source in the chromosphere is unlikely to be
monochromatic, and so this situation of partial transmission and partial
reflection with the reduced FIP fractionation will be ubiquitous in the
solar atmosphere. This does not agree with observations, for which chromosphere
``A'' is a much better match. We therefore argue that if the FIP effect is due
to the ponderomotive force of Alfv\'en waves in the chromosphere, these
must have a source in the corona. We return to this thought in subsection 4.1.

Although this calculation has been done for a closed loop, we expect
that this chromospheric wave pattern will also arise at the
footpoint of an open field line in a coronal hole. In fact the
character of our chromospheric solution matches well with that found
in the open field case by \citet{cranmer05}, subsequently shown in
\citet{cranmer07} to exhibit FIP fractionation similar to that
observed in the fast solar wind. We emphasize this point by showing
in Figures 14 and 15 the coronal and chromospheric portions of an
open field flux tube. In this case we start the integration at an
altitude of $5\times 10^5$ km with purely outgoing waves, and work
back to the solar surface. This restricts us to the region where the
solar wind outflow speed is still much lower than the Alfv\'en
speed, in keeping with our assumption of $u << V_A$ above. We take
magnetic field from \citet{banaskiewicz98}, modified by
\citet{cranmer05}, and choose a density scale height to match the
observed and modeled density profiles in \citet{laming04b}. Figure
14 shows $\delta v$ and $\delta B/\sqrt{4\pi\rho}$, chosen to match
the observational and modeling constraints in \citet{cranmer05}.
Figure 15 shows the extension of these variables into chromosphere
``B'', in a similar manner to the previous figures. While there is
much more to be said about the wave properties in open field lines,
the important point to be made here is that the ponderomotive force
naturally produces a very small fractionation in this geometry. This
is consistent with observed abundances in the fast solar wind
\citep{zurbuchen02} and in coronal holes \citep{feldman98}. A
tabulation of coronal hole fractionations is given in Table 2, in a
similar format to that in Table 1, using the coronal hole incident
spectrum of \citet{vernazza78}.

\section{Discussion}
\subsection{Alfv\'en Wave Energy Fluxes}
It appears from the forgoing that the abundance anomalies observed in various
regions of the solar corona may yield inferences on the energy fluxes of
Alfv\'en waves in the chromosphere. Our initial considerations then imply that
wave energy fluxes sufficient to heat the solar corona or accelerate the solar wind
are necessary to produce the correct fractionation. Energy fluxes
observed in slow mode and fast mode waves are not sufficient to heat the
solar corona \citep{erdelyi07}. The detection of Alfv\'en waves, the favored mode
for transporting energy to the solar corona, is much harder, since they are
incompressible and can only be revealed through Doppler shifts or motions, which
become hard to see in inhomogeneous conditions where Alfv\'en waves on neighboring
flux surfaces can propagate at different speeds and lose phase coherence. However
\citet{tomczyk07} claim the detection of Alfv\'en waves in the solar corona,
albeit with insufficient energy flux to heat the corona. \citet{vandoors08}
argue that the detected waves are in fact kink mode waves, for the reasons suggested
above. \citet{depontieu07} observe transverse waves in the chromosphere, which they
argue should be interpreted as Alfv\'en waves in the absence of a chromospheric
waveguide. However these waves are inferred from the observed oscillations of
spicules, which clearly have radial structure. The energy flux detected by these
authors $\sim 10^5$ ergs cm$^{-2}$ s$^{-1}$
is close to being sufficient to heat the solar corona or accelerate the solar wind.

In this paper we argue that the FIP effect is due to the ponderomotive force associated
with transverse waves in the chromosphere. Longitudinal MHD waves do not generate
electric field. In order to generate the observed FIP fractionation, the energy
fluxes associated with these waves need to be of order $10^6 - 10^7$
ergs cm$^{-2}$ s$^{-1}$, much closer to those required for coronal heating. It
is clear that FIP fractionation is associated with the transmission of waves between
the chromosphere and the corona, and correspondingly we argue that the required
transverse waves should be identified as Alfv\'en waves to meet this condition.
The fast mode totally internally
reflects somewhere in the transition region or low corona (Schwartz
\& Leroy 1982, Leroy \& Schwartz 1982).

The nonthermal mass motions predicted in the coronal section of this loop are
higher than observed. In the transition region, this is not necessarily a problem
since ample evidence exists to show that the ``classical'' transition regions
of coronal loops are rarely observed, being masked as they are by a population
of smaller ``unresolved fine structures'' \citep{feldman83,feldman87,peter01}. In the
corona, this may also be true if the heating occurs in thin filament or shells
as in Alfv\'en resonance models \citep[e.g.][]{terradas08} while the rest of the
emitting loop undergoes much slower oscillations. Another possibility might be
the generation of turbulence following coronal reconnection events associated with
nanoflares \citep{dahlburg05}. In each case it is likely that the turbulence would
actually be produced in the coronal section of the loop, not in the chromosphere, and
will also be in resonance with the loop. This also appears to be the conclusion to be
drawn from section 3. Chromosphere ``A'' where waves leak down from the corona
before being reflected back again gives stronger and more consistent FIP
fractionations for a wide variety of wave spectra than chromosphere ``B'', where
waves are incident upwards on the loop from the chromosphere below.
We therefore argue that the FIP
effect is more likely to arise with a coronal source of Alfv\'en waves, rather
than a chromospheric source as originally conceived in \citet{laming04}, and that
this inference will constrain the means by which the corona may be heated.

\subsection{Helium and Neon in the Solar Corona}
The fractionations computed in this paper differ from those in \citet{laming04}
in three notable ways. First, as the turbulent energy density increases, the
fractionation does not increase without limit but saturates at values
broadly consistent with those observed. This is due to a refinement in our
formalism discussed above, where the wave oscillation velocity is included in
the ion and neutral partial pressures in equation 3.

The second is that at high turbulence levels, He becomes
significantly depleted relative to H. Comparing Tables 1 and 2, we
find a stronger depletion in the coronal loop, representative of the
slow speed solar wind, than we would in a coronal hole, the source
of the fast wind. The abundance ratio He/H in the fast solar wind is
fairly constant at about 5\% \citep{aellig01,zurbuchen02}, or a
depletion of 0.59 from the photospheric value of 8.5\%. He/H in the
slow speed solar is lower, and generally more variable.
\citet{aellig01} and \citet{kasper07} find He/H varying with wind
speed, with these variations being more pronounced at solar minimum,
where He/H $\sim 1\%$ for speeds below 300 km s$^{-1}$, approaching
4.5\% for speeds above 500 km s$^{-1}$. At solar maximum, He/H is
always in the range 3.5 - 5\%. \citet{kasper07} also find a
dependence on heliographic latitude during periods of solar minimum,
with lower He/H being found closer to the heliographic equator.
Table 1 gives values of He/H down to about 3.5\%. Overall though,
our modeled values for the abundance ratio He/H are very
encouragingly consistent with observations, lending confidence to
our approach.

The third, and most controversial new feature is the similar depletion predicted
for Ne. This was originally suggested by \citet{drake05} from a survey of the
Ne/O abundance ratio in a sample of late-type stellar coronae, as a solution to the
problem in helioseismology presented by the reduction in the solar photospheric
abundance of O \citep[see][and references therein]{caffau08}. Specifically, \citep{basu04}
the depth
of the solar convection zone demands a metallicity higher than that coming
from the standard solar composition, with the O abundance revised
downwards by nearly a factor of 1.5 \citep{asplund04} from \citet{grevesse98}.
Ne, having no photospheric lines on which to base an abundance
measurement, was suggested as the element most likely to resolve this by having
a higher postulated abundance \citep[e.g.][]{bahcall05,basu08}. \citet{drake05}
find coronal Ne/O typically $\sim 0.4$ in stars which exhibit either no FIP effect or an
inverse FIP effect, and argued that the general consistency of Ne/O among their
sample of 21 stars suggests no significant fractionation between Ne and O here between
photosphere and corona. The solar coronal abundance ratio Ne/O, measured at
$\simeq 0.15-0.18$ \citep{schmelz05,young05} would imply therefore that Ne is depleted in
the solar corona relative to the photosphere, similarly
to He. Our calculations in Table 1 provide some support to this view, especially at higher
turbulence levels, where Ne/O is about 0.5 of its photospheric value.

\subsection{Fractionation in the Low Chromosphere}
One main feature of Laming (2004) model and the
calculations presented above is that the fractionation is
predicted to occur relatively high up in the chromosphere, at altitudes
greater than 2000 km. However in the literature there are already
indications that, at least in active regions and flares, that
fractionation should set in lower down.

In an analysis of HRTS II (the Naval Research Laboratory's High
Resolution Telescope and Spectrograph) data, Athay (1994) observed
variations in the C I 1561\AA\ /Fe II 1563 \AA\ line intensity
ratio. Compared to plage regions around a sunspot, the sunspot
itself has a higher ratio  C I/Fe II, while surrounding C I dark
flocculi have a lower ratio. Similar results are found by Doschek,
Dere \& Lund (1991) and Feldman, Widing \& Lund (1990). This absence
of fractionation in the sunspot presumably relates to the absence of
acoustic waves in sunspots \citep[e.g.][]{muglach05} because
convection is inhibited by the strong magnetic field
\citep{parchevsky07}. The fact that this is observed in lines of C I
and Fe II suggests that fractionation must set in at lower altitudes
than originally modeled by Laming (2004; see Figure 1 (left and
right panels), where O and C are becoming ionized in the region of
fractionation, and one would expect neutral O I and C I to be
emitted from lower, unfractionated layers).

The existence of fractionation at these low altitudes also offers a
possible explanation of the observation by Phillips et al. (1994)
who found rather small difference in the abundances of Fe determined
from soft X-ray flare plasma, compared with that lower down in the
atmosphere, determined from the Fe K$\beta$ fluorescent line, rather
than the strong FIP effect expected. More recently Murphy \& Share
(2005) studied $\gamma$-ray emission from flares. Protons
accelerated into the chromosphere by the flare excite $\gamma$-ray
emission from the ambient plasma when its density reaches about
$10^{14}-10^{15}$ cm$^{-3}$. Element abundances determined from the
resulting $\gamma$-ray spectrum show the presence of a FIP
fractionation. The densities at which this occurs correspond to the
low chromosphere where sound and Alfv\'en speeds are approximately
equal, and certainly not the Lyman $\alpha$ plateau region where
fractionation is expected in the Laming (2004) model. A search for
FIP fractionation in photospheric lines i.e. below the chromosphere
(Sheminova \& Solanki 1999) reveals very little, if any
fractionation. Thus all available observational evidence suggests
that the low chromosphere as another plausible place for FIP
fractionation to occur.

We speculate that the growth of Alfv\'en waves from
sound waves near the $\beta =1$ layer will give an extra
ponderomotive force in this region that can account for this.
Zaqarashvili \& Roberts (2006) give a treatment of the parametric
conversion of sound waves into Alfv\'en waves which requires
$\beta=1$ when both are traveling in the same direction along the
magnetic field. This distinguishes it from the phenomenon of
mode conversion, which requires nonzero wavevector perpendicular to
the magnetic field to proceed (e.g. McDougall \& Hood 2007), and
parametric conversion lower down in high $\beta$ plasma where the
sound waves must be oblique \citep{zaqarashvili02}.
Waves impinging on the chromosphere from below are fast
magnetoacoustic waves from the high $\beta$ (gas pressure/magnetic
pressure) solar interior. In the absence of mode conversion, these
retain their acoustic character propagating as a slow mode wave when
$\beta << 1$ further up in the chromosphere (McDougall \& Hood
2007). At the altitude where $\beta\simeq 1$ (i.e. where the phase
speeds of magnetic and acoustic waves are similar) these waves can
mode convert into other MHD wave modes (Bogdan et al. 2003). This
would be consistent with the findings of Sheminova \& Solanki
(1999), who find essentially no FIP effect at photospheric
altitudes. Acoustic waves will produce no ponderomotive force, and
only once mode conversion to the other MHD modes has occurred can
fractionation proceed.

\section{Conclusions}
In conclusion then we have refined the model of \citet{laming04} for FIP
fractionations arising from the ponderomotive force as Alfv\'en waves
propagate through the chromosphere. We have implemented a non-WKB treatment
of the wave transport, which can be further modified to include the effects
of wave growth and damping, and made a correction to the previous formalism
to include the Alfv\'en wave transverse velocity in the chromospheric ion and
neutral partial pressures. The new effects are a saturation of the FIP effect
at the correct level, and predicted depletions in the coronal abundances of
He and Ne, again consistent with observations. We find the best match to the
observed coronal or solar wind element abundances arises for models with an
Alfv\'en wave energy fluxes sufficient to heat the corona or accelerate
the solar wind. The inference that a coronal source of Alfv\'en waves provides
a FIP effect better matching the observations suggests that coronal abundance
anomalies may provide novel insights into the coronal heating mechanism(s).

\acknowledgements
This work was supported by NASA Contract NNG05HL39I, and by basic research funds of
the Office of Naval Research. I thank Daniel Savin, Cara Rakowski and an anonymous
referee for comments on the manuscript.

\appendix
\section{The Ponderomotive Force}
The ponderomotive force arises from the effects of wave refraction
in an inhomogeneous plasma. In a nonmagnetic plasma, the refractive
index, $\sqrt{\epsilon}$, is given by $\epsilon=1-\omega _p^2/\omega
^2$ where $\omega _p$ is the plasma frequency. Waves are refracted
to high refractive index, which means low plasma density. The
increased wave pressure can then expel even more plasma from the low
density region, leading to ducting instabilities. In magnetic plasma, $\epsilon=1-\omega
_p^2/\left(\omega ^2-\Omega ^2\right)$, where $\Omega$ is the ion
cyclotron frequency. Thus waves refract to high density regions, and
plasma is attracted to regions of high wave energy density. A simple
expression for the ponderomotive force on an ion may be derived as
follows. The Lagrangian density for a system of thermal plasma of
density $n$ with particle mass $m$ and waves is
\begin{equation}
L=\sum _i{1\over 2}m_i\left(v_{thi,i}^2+v_{osc,i}^2\right)
+\sum _i {q_i\over c}\left({\bf v}_{th,i}+{\bf v}_{osc,i}\right)\cdot
\delta {\bf A} +
{\epsilon\delta E^2-\delta B^2\over 8\pi}
\end{equation}
where $v_{th,i}$ is the thermal speed and $v_{osc,i}$ is the
oscillatory speed induced by the wave of particle $i$, with mass $m_i$,
and charge $q_i$. Wave electric and magnetic fields are given by $\delta {\bf E}$ and
$\delta {\bf B}$ respectively, and $\delta {\bf A}$ is the vector potential. We have omitted the interaction term involving the
electrostatic potential, since this is constant in a neutral plasma.
Putting
$\delta B^2/8\pi = \sum _imv_{osc,i}^2/2 +\delta E^2/8\pi$ and ${\bf v}_{osc,i}\cdot
\delta {\bf A}=0$ for MHD waves, then
\begin{equation}
L=\sum _i{1\over2}mv_{thi,i}^2
+\sum _i {q_i\over c}{\bf v}_{th,i}\cdot{\bf A}
+{\left(\epsilon-1\right)\delta E^2\over
8\pi}=\sum _i{1\over2}mv_{thi,i}^2+
\sum _i {q_i\over c}{\bf v}_{th,i}\cdot{\bf A}+\sum
_i{q_i^2\over 2m_i\left(\Omega _i^2-\omega ^2\right)}{\delta E^2}.
\end{equation}
The ``$z$'' Euler-Lagrange equation gives
\begin{equation}
{d\over dt}\left(mv_{th,iz}\right)= {q_i^2\over 2m_i
\left(\Omega _i^2-\omega ^2\right)}{d\delta E^2\over dz},
\end{equation}
neglecting the spatial variation of $B$ and hence $\Omega _i$, and
evaluating for the component of $v_{th,i}$ orthogonal to ${\bf A}$ and
${\bf B}$. This is
the same as the expression derived by Landau, Lifshitz \& Pitaevskii
(1984), and agrees with earlier work (e.g. Lee \& Parks 1983) if
$\delta E^2=\delta E_p^2/2$, where $\delta E_p$ is the peak electric field
in the wave, giving a ponderomotive force
\begin{equation}
F_i={q_i^2\over 4m_i\left(\Omega _i^2-\omega ^2\right)}{d\delta E_p^2\over
dz}.
\end{equation}
When $\omega << \Omega _i$, the ponderomotive acceleration is thus {\em
independent of ion mass}, which is one crucial property relevant to
obtaining an almost mass independent fractionation as observed. It
is also independent of ion change, so long as the ion is charged
(and not neutral). Litwin \& Rosner (1998) give a similar expression
derived from the ${\bf j}\times {\bf B}$ term in the MHD momentum
equation.

\section{The Non-WKB Transport Equations}
We start from the linearized MHD force and induction equations,
\begin{equation}
\rho{\partial\delta {\bf v}\over\partial t}+ \nabla\left(\rho{\bf
u}\cdot\delta {\bf v}\right)={\left(\nabla\times\delta{\bf
B}\right)\times{\bf B}\over 4\pi}={\left({\bf
B}\cdot\nabla\right)\delta {\bf B}-\left(\nabla\delta{\bf
B}\right)\cdot {\bf B}\over 4\pi},
\end{equation}
and
\begin{equation}
{\partial\delta {\bf B}\over\partial t}=\nabla\times\left(\delta
{\bf v}\times {\bf B}\right) +\nabla\times\left({\bf u}\times\delta
{\bf B}\right) = \left({\bf B}\cdot\nabla\right)\delta {\bf
v}-\delta {\bf B}\nabla\cdot {\bf u}-\left({\bf
u}\cdot\nabla\right)\delta {\bf B},
\end{equation}
where ${\bf u}$ and ${\bf B}$ are the unperturbed velocity and
magnetic field, $\delta {\bf v}$ and $\delta {\bf B}$ are the
perturbations, and $\rho$ is the density.
Equation (B1) is rewritten using $\nabla\left(\rho
{\bf u}\cdot\delta {\bf v}\right)=\rho {\bf
u}\times\nabla\times\delta {\bf v}+\delta {\bf
v}\times\nabla\times\left(\rho {\bf u}\right)+\left(\rho {\bf
u}\cdot\nabla\right)\delta {\bf v}+\left(\delta {\bf
v}\cdot\nabla\right)\rho {\bf u}$ to yield
\begin{equation}
{\partial\delta {\bf v}\over\partial t}+\left({\bf
u}\cdot\nabla\right)\delta {\bf v}={\bf V}_A\cdot\nabla\left(\delta
{\bf B}\over\sqrt{4\pi\rho}\right)+{\delta {\bf
B}\over\sqrt{4\pi\rho}}{{\bf V}_A\cdot\nabla\rho\over 2\rho}+
{\left(\nabla {\bf B}\right)\cdot\delta {\bf B}\over
4\pi\rho}-{\delta {\bf v}\cdot\nabla\left(\rho {\bf
u}\right)\over\rho}
\end{equation}
where ${\bf V}_A={\bf B}/\sqrt{4\pi\rho}$ is the Alfv\'en velocity.
Writing $\left(\nabla {\bf B}\right)\cdot\delta {\bf
B}=\left(\partial B_x/\partial x\right)\delta {\bf
B}=-\left(\partial B_z/\partial z\right)\delta {\bf B}/2$ since
$\nabla\cdot {\bf B}=0$ (assuming $\partial B_x/\partial x=
\partial B_y/\partial y$), and similarly for $\left(\nabla\rho {\bf
u}\right)\cdot\delta {\bf v}$, and using $\partial\left(\rho
u_z/B_z\right)/\partial z =0$ gives
\begin{equation}
{\partial\delta {\bf v}\over\partial t}+\left({\bf
u}\cdot\nabla\right)\delta {\bf v}={\bf V}_A\cdot\nabla\left(\delta
{\bf B}\over\sqrt{4\pi\rho}\right)+{\delta {\bf
B}\over\sqrt{4\pi\rho}}{V_A\over 2H_D}-{\delta {\bf
B}\over\sqrt{4\pi\rho}}{V_A\over 2H_B}+\delta {\bf v}{u\over 2H_B}.
\end{equation}
Here $1/H_B=\partial\ln B_z/\partial z$, $1/H_D=\partial\ln\rho
/\partial z$, and below $1/H_A=\partial\ln V_A/\partial z$. Similar
manipulations give the induction equation in the form
\begin{equation}
{\partial\over\partial t}\left(\delta {\bf
B}\over\sqrt{4\pi\rho}\right)+\left({\bf u}\cdot\nabla\right){\delta
{\bf B}\over\sqrt{4\pi\rho}}=\left({\bf V}_A\cdot\nabla\right)\delta
{\bf v}+{\delta {\bf B}\over\sqrt{4\pi\rho}}{u\over 2H_D}+\delta
{\bf v}{V_A\over 2H_B}-{\delta {\bf B}\over\sqrt{4\pi\rho}}{u\over
2H_B}.
\end{equation}
Taking equation (B4) plus or minus equation (B5) and rearranging
gives the final result,
\begin{equation} {\partial I_{\pm}\over\partial t}+\left(u\pm
V_A\right){\partial I_{\pm}\over\partial z}= \left(u\pm
V_A\right)\left({I_{\pm}\over 4H_D}+{I_{\mp}\over 2H_A}\right),
\end{equation}
where $I_{\pm}=\delta {\bf v}\pm \delta {\bf B}/\sqrt{4\pi\rho}$,
representing waves propagating in the $\mp$ z-directions.

\clearpage
\begin{table}
\begin{center}
\caption{Coronal FIP Fractionations}
\begin{tabular}{l|rrrrrrr|rrr}
\tableline\tableline
ratio & \multicolumn{7}{c}{relative wave energy flux}& \multicolumn{2}{c}{obs.}\\
 & 1/64& 1/16& 1/4& 1& 4& 16& 64& a& b& c\\
\tableline
He/H & 1.0& 1.1& 1.1& 0.79& 0.55& 0.46& 0.44& 0.68\\
C/H & 1.1& 1.3& 1.5& 1.2& 0.84& 0.66& 0.60& 1.36\\
N/H & 1.1& 1.3& 1.5& 1.2& 0.86& 0.72& 0.69& 0.72\\
O/H & 1.1& 1.4& 1.9& 1.8& 1.3& 1.1& 1.1& 1.00\\
Ne/H& 1.1& 1.2& 1.4& 1.1& 0.76& 0.63& 0.60& 0.58\\
Na/H& 1.2& 1.9& 4.6& 9.8& 13.& 13.& 13.& & 7.8${+13\atop -5}$&
2.1${+2\atop -1}$\\
Mg/H& 1.2& 1.7& 3.5& 6.1& 6.9& 6.6& 6.3& 2.58& 2.8${+2.3\atop
-1.3}$&
3.0${+1.7\atop -1.1}$\\
Al/H& 1.2& 1.6& 2.8& 3.6& 3.1& 2.7& 2.5& & 3.6${+1.7\atop -1.2}$&
6.8${+4.0\atop -2.5}$\\
Si/H& 1.2& 1.7& 3.0& 4.2& 3.8& 3.2& 3.1& 2.49& 5.1${+3\atop -1.9}$\\
S/H & 1.1& 1.4& 2.0& 1.9& 1.4& 1.1& 1.0& 1.62& 2.2$\pm 0.2$&
2.3${+1.3\atop -0.8}$\\
Ar/H& 1.1& 1.4& 1.8& 1.5& 1.1& 0.91& 0.87\\
K/H & 1.3& 2.3& 8.0& 25.& 38.& 41.& 42.& & 1.8${+0.4\atop -0.6}$&
4.2${+6.3\atop -2.5}$\\
Ca/H& 1.2& 1.6& 2.6& 3.0& 2.3& 1.9& 1.8& & 3.5${+4.3\atop -1.9}$&
3.1${+1.8\atop -1.1}$\\
Fe/H& 1.2& 1.6& 2.8& 3.3& 2.6& 2.2& 2.1& 2.28& 4.4$\pm 0.5$\\
Ni/H& 1.2& 1.6& 2.4& 2.5& 1.9& 1.5& 1.5\\
Kr/H& 1.1& 1.4& 1.9& 1.7& 1.3& 1.1& 1.0\\
Rb/H& 1.3& 2.3& 7.4& 19.& 25.& 25.& 25.\\
W/H & 1.3& 2.3& 6.9& 16.& 18.& 17.& 17.\\
\tableline
\end{tabular}
\end{center}
\tablecomments{FIP fractionations corresponding to the chromospheric model in
Figure 11, for relative wave energy flux = 1. Other relative wave energy
fluxes are given to show the behavior of the FIP fractionation. Observational
ratios are taken from, (a) Zurbuchen et al. (2002), relative to O,
(b) Bryans et al. (2008), relative to the mean of O, Ne and Ar, and (c)
Giammanco et al. (2008), relative to H.}
\label{tab1}
\end{table}

\begin{table}
\begin{center}
\caption{Coronal Hole FIP Fractionations}
\begin{tabular}{l|rrrrrrr|r}
\tableline\tableline
ratio & \multicolumn{7}{c}{relative wave energy flux}& obs.\\
 & 1/64& 1/16& 1/4& 1& 4& 16& 64\\
\tableline
He/H& 1.0& 1.0& 1.0& 1.0& 0.95& 0.92& 0.91& 0.58\\
C/H & 1.0& 1.0& 1.1& 1.1& 0.99& 0.96& 0.95& 1.41\\
N/H & 1.0& 1.0& 1.1& 1.1& 1.0& 0.97& 0.96& 0.93\\
O/H & 1.0& 1.1& 1.1& 1.2& 1.1& 1.0& 1.0& 1.00\\
Ne/H& 1.0& 1.0& 1.1& 1.0& 0.98& 0.94& 0.93& 0.47\\
Na/H& 1.2& 1.3& 1.4& 1.4& 1.3& 1.2& 1.2\\
Mg/H& 1.1& 1.2& 1.3& 1.2& 1.2& 1.1& 1.1& 1.92\\
Al/H& 1.1& 1.1& 1.1& 1.1& 1.1& 1.0& 1.0\\
Si/H& 1.1& 1.1& 1.1& 1.1& 1.1& 1.0& 1.0& 1.86\\
S/H & 1.0& 1.0& 1.1& 1.1& 1.0& 0.98& 0.97& 1.56\\
Ar/H& 1.0& 1.1& 1.1& 1.1& 1.0& 1.0& 0.99\\
K/H & 1.2& 1.2& 1.3& 1.3& 1.2& 1.2& 1.2\\
Ca/H& 1.0& 1.1& 1.1& 1.1& 1.1& 1.0& 1.0\\
Fe/H& 1.0& 1.1& 1.1& 1.1& 1.1& 1.0& 1.0& 1.67\\
Ni/H& 1.0& 1.1& 1.1& 1.1& 1.1& 1.0& 1.0\\
Kr/H& 1.0& 1.1& 1.1& 1.1& 1.1& 1.0& 0.99\\
Rb/H& 1.2& 1.2& 1.3& 1.3& 1.3& 1.2& 1.2\\
W/H & 1.1& 1.1& 1.2& 1.2& 1.1& 1.1& 1.1\\
 \tableline
\end{tabular}
\end{center}
\tablecomments{FIP fractionations corresponding to the chromospheric model in
Figure 15, for relative wave energy flux = 1. Other relative wave energy
fluxes are given to show the behavior of the FIP fractionation. Observational results
are taken from Zurbuchen et al (2002), given relative to O.}
\label{tab2}
\end{table}

\clearpage

\begin{figure}[t]
\epsscale{1.0} \plotone{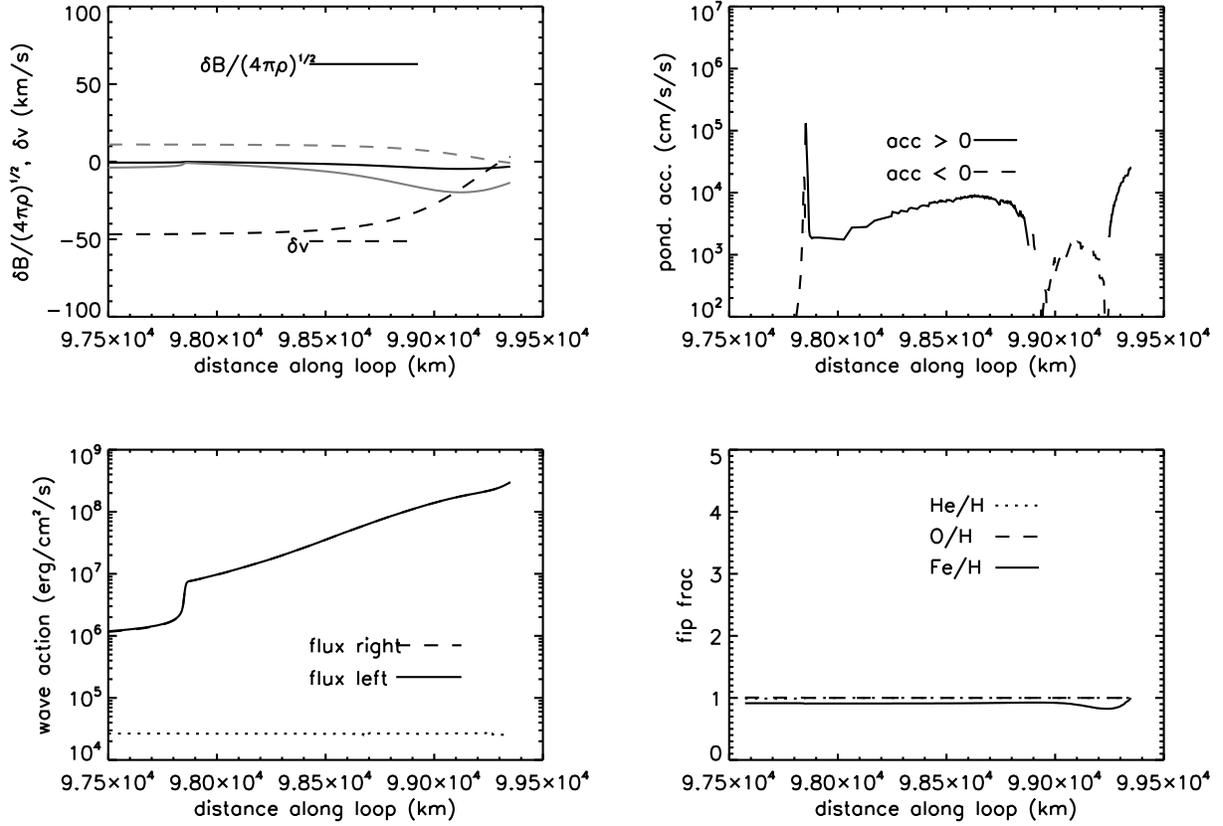} \figcaption[f11.ps]{Same as figure 8
(off resonance case)
for the right hand side chromosphere ``B'', where Alfv\'en waves are
launched up from the convection zone. The wave energy flux has been
increased by a factor 20, leading to stronger coronal nonthermal
motions, and stronger fractionation. Waves are now reflected, and a
small inverse FIP effect results.\label{fig11}}
\end{figure}

\begin{figure}[t]
\epsscale{1.0} \plotone{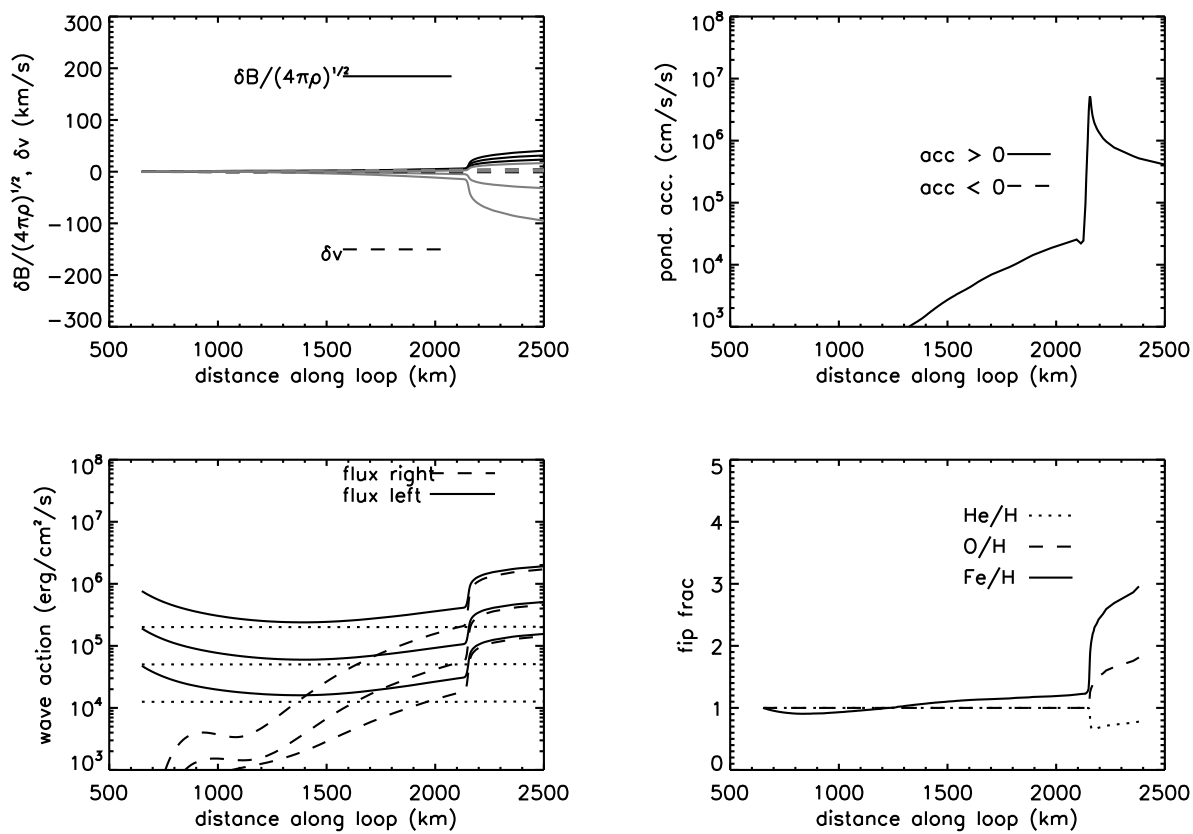} \figcaption[f12.ps]{Same as figure 9
giving the first three panels for the left hand chromosphere ``A'', where
waves leak down from the corona. The extra bottom right panel shows
the FIP fractionation for Fe, O, and He, relative to H. Three wave
frequencies are now introduced to simulate more nearly a realistic
chromospheric power spectrum. \label{fig12}}
\end{figure}

\begin{figure}[t]
\epsscale{1.0} \plotone{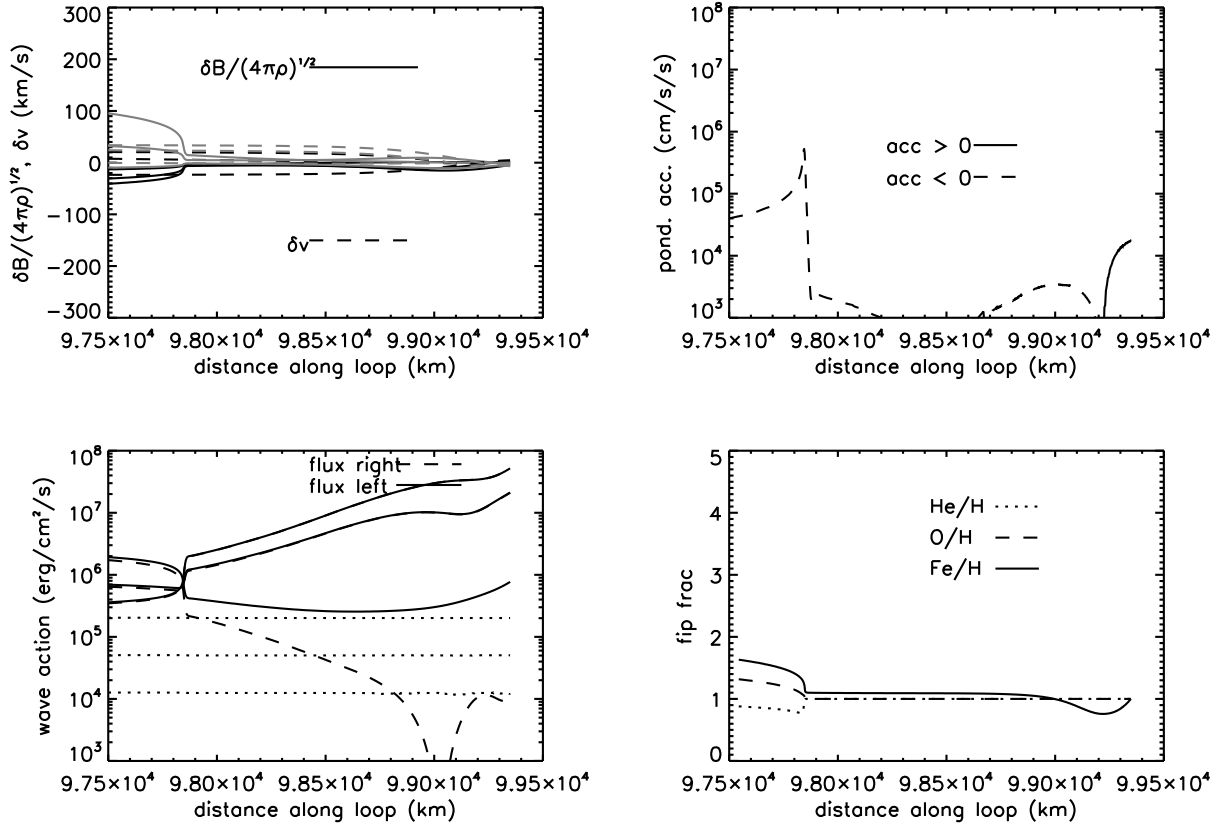} \figcaption[f13.ps]{Same as figure
11 for the right hand side chromosphere ``B'', where three Alfv\'en waves are
launched up from the convection zone. The FIP fractions are reduced from
those with a single incident wave, even though one of the waves here is on
resonance. The contributions to the partial pressure of the other waves
``dilute'' the fractionation, by increasing the value of $v_s^2$ in the
denominator of the integral in equation 6, without increasing the ponderomotive
acceleration.\label{fig13}}
\end{figure}

\begin{figure}[t]
\epsscale{1.0} \plotone{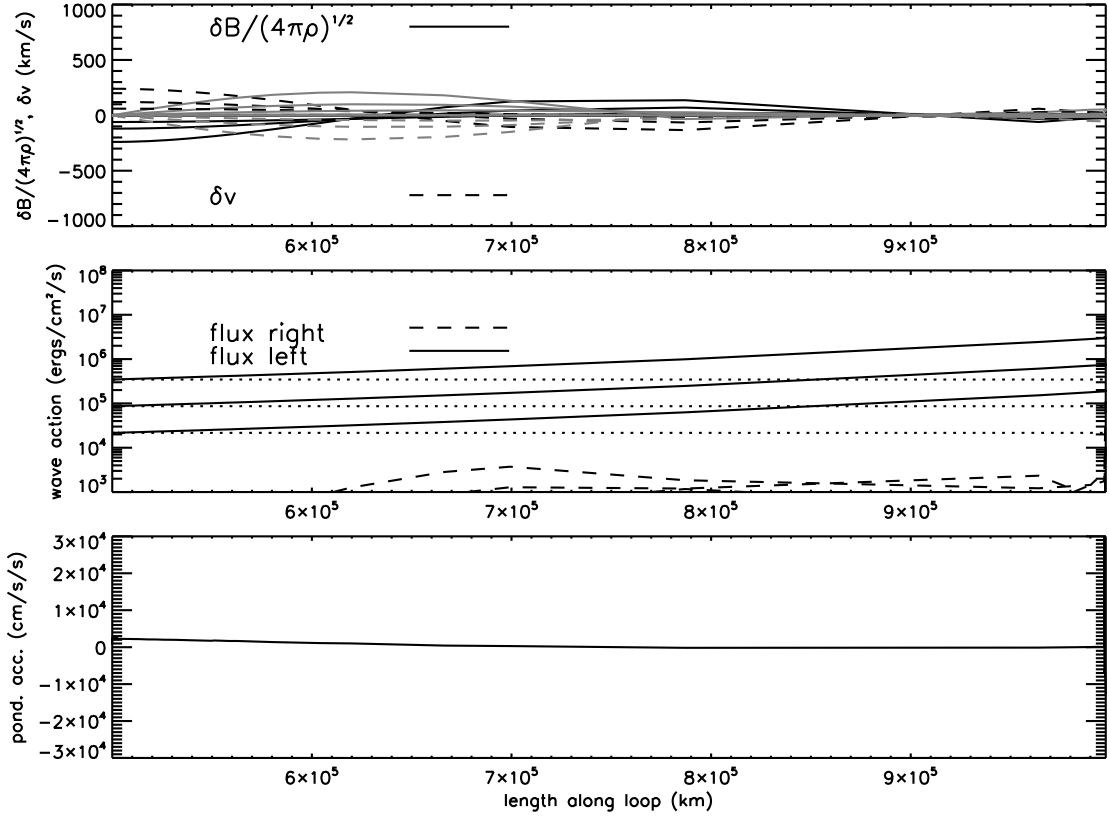} \figcaption[f12.ps]{Coronal section
of open field region up to 500,000 km altitude, showing from top:
Els\"asser variables in km s$^{-1}$ ($\delta B/\sqrt{4\pi\rho}$
solid lines, $\delta v$ dashed lines), with black lines for real
parts and gray lines for imaginary parts. Middle; wave energy fluxes
in ergs cm$^{-2}$ s$^{-1}$. Bottom, the ponderomotive acceleration
in cm s$^{-2}$. \label{fig14}}
\end{figure}

\begin{figure}[t]
\epsscale{1.0} \plotone{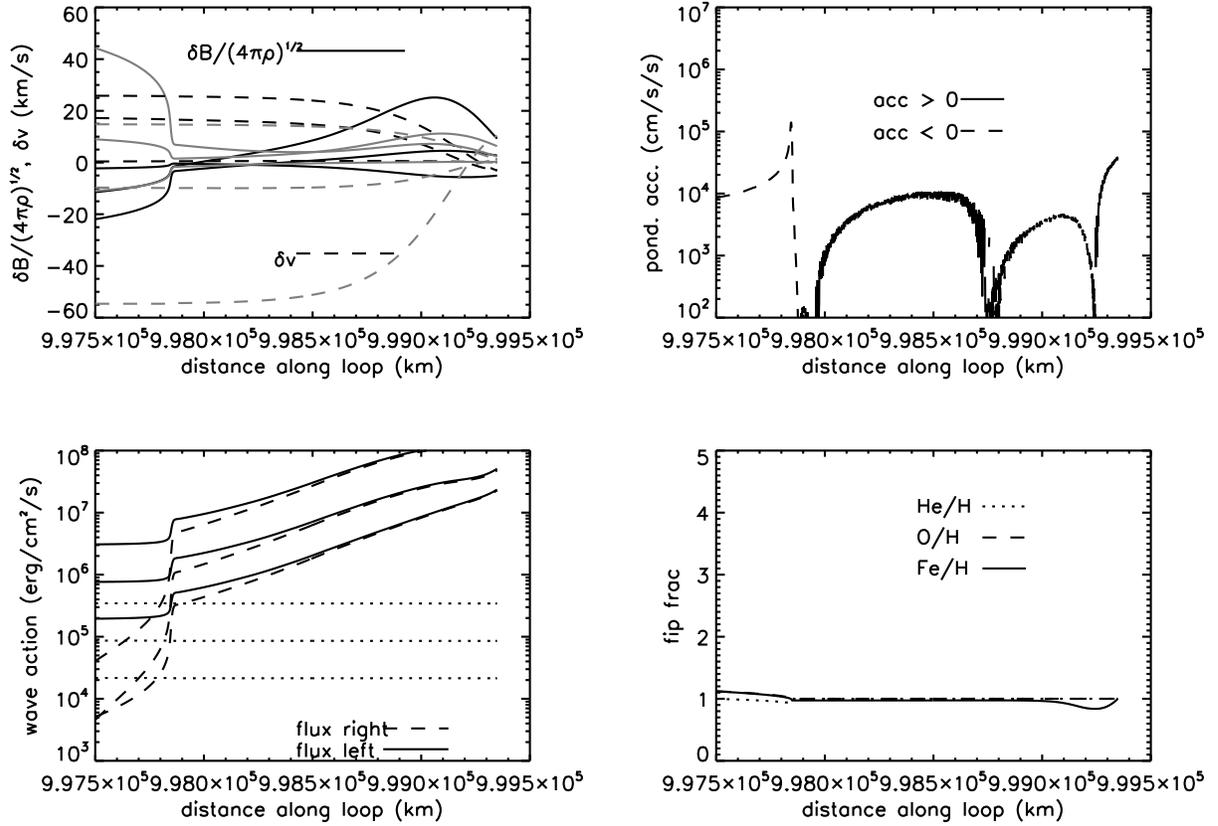} \figcaption[f13.ps]{Same as figure
11 for the right hand side chromosphere ``B'', where Alfv\'en waves are
launched up from the convection zone into an open field region.
The FIP fractionations are
evaluated with an incident coronal hole spectrum, as opposed to that
for an active region, and show the absence of strong fractionation consistent
with observations of the fast solar wind and coronal holes.\label{fig15}}
\end{figure}
\end{document}